\documentclass[journal,twoside,web]{ieeecolor}
\usepackage{jsen}
\usepackage{cite}
\usepackage{amsmath,amssymb,amsfonts}
\usepackage{algorithm}
\usepackage{algpseudocode}
\usepackage{graphicx}
\usepackage{textcomp}
\usepackage{wrapfig}
\usepackage{caption}
\usepackage{multirow}
\usepackage{makecell}
\usepackage{hyperref}
\usepackage{subcaption}

\def\BibTeX{{\rm B\kern-.05em{\sc i\kern-.025em b}\kern-.08em
    T\kern-.1667em\lower.7ex\hbox{E}\kern-.125emX}}
\definecolor{abstractbg}{rgb}{0.89804,0.94510,0.83137}
\begin{document}
\title{Joint Sparse Representations and Coupled Dictionary Learning in Multi-Source Heterogeneous Image Pseudo-color Fusion}
\author{Long Bai, \IEEEmembership{Graduate Student Member, IEEE}, Shilong Yao, Kun Gao, Yanjun Huang, Ruijie Tang, \\ Hong Yan, \IEEEmembership{Fellow, IEEE}, Max Q.-H. Meng, \IEEEmembership{Fellow, IEEE}, Hongliang Ren, \IEEEmembership{Senior Member, IEEE}
\thanks{Manuscript received on August 28, 2023, accepted on October 08, 2023.
The work was supported by the National Natural Science Foundation of China (Grant U2241275 and 61827814); Beijing Natural Science Foundation (Grant Z190018); China High-resolution Earth Observation System Project (Grant 52-L10D01-0613-20/22);
Hong Kong RGC CRF C4026-21GF, CRF C4063-18G, GRF 14203323, GRF 14216022, GRF 14211420, GRS 3110167, NSFC/RGC Joint Research Scheme N\_CUHK420/22; Shenzhen-Hong Kong-Macau Technology Research Programme (Type C) 202108233000303; GBABF \#2021B1515120035;
City University of Hong Kong \#11204821. \{Corresponding to: H. Ren, K. Gao.\}}
\thanks{L. Bai, R. Tang, and H. Ren are with the Department of Electronic Engineering, The Chinese University of Hong Kong, Hong Kong 999077, China. (e-mail: b.long@ieee.org, ruijie.tang@link.cuhk.edu.hk)}
\thanks{S. Yao and H. Yan are with the Department of Electrical Engineering, City University of Hong Kong, Hong Kong 999077, China. (e-mail: shilong.yao@my.cityu.edu.hk, h.yan@cityu.edu.hk)}
\thanks{S. Yao and M. Meng are with Shenzhen Key Laboratory of Robotics Perception and Intelligence, Shenzhen, China, and the Department of Electronic and Electrical Engineering, Southern University of Science and Technology, Shenzhen 518055, China. (e-mail: max.meng@ieee.org)}
\thanks{K. Gao and Y. Huang are with the Key Laboratory of Photoelectronic Imaging Technology and System, Ministry of Education of China, Beijing Institute of Technology, Beijing 100081, China.
        (e-mail: \{gaokun, 3220210458\}@bit.edu.cn)
        }
\thanks{H. Ren is also with the Shun Hing Institute of Advanced Engineering, The Chinese University of Hong Kong, Hong Kong 999077, China; and with the Department of Biomedical Engineering, National University of Singapore (NUS), Singapore 117575, Singapore, and NUS (Suzhou) Research Institute, Suzhou 215123, China. (e-mail: hlren@ieee.org)}
}
\IEEEtitleabstractindextext{%
\fcolorbox{abstractbg}{abstractbg}{%
\begin{minipage}{\textwidth}%
\begin{wrapfigure}[12]{r}{2.85in}%
\includegraphics[width=2.7in]{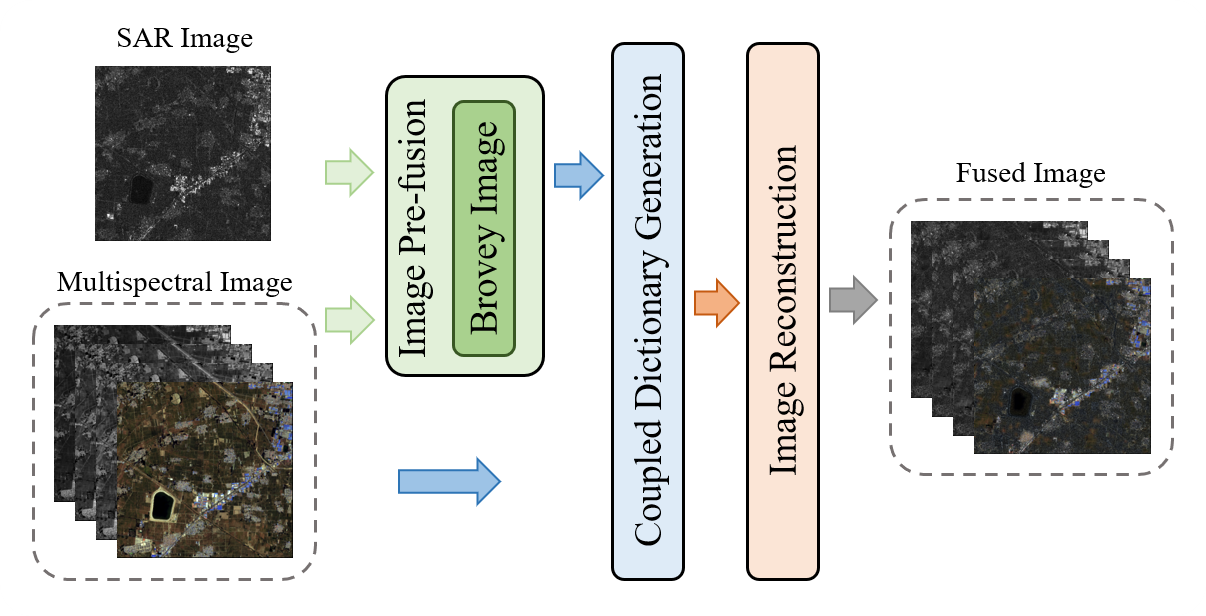}%
\end{wrapfigure}%

\begin{abstract}
Considering that Coupled Dictionary Learning (CDL) method can obtain a reasonable linear mathematical relationship between resource images, we propose a novel CDL-based Synthetic Aperture Radar (SAR) and multispectral pseudo-color fusion method. Firstly, the traditional Brovey transform is employed as a pre-processing method on the paired SAR and multispectral images. Then, CDL is used to capture the correlation between the pre-processed image pairs based on the dictionaries generated from the source images via enforced joint sparse coding. Afterward, the joint sparse representation in the pair of dictionaries is utilized to construct an image mask via calculating the reconstruction errors, and therefore generate the final fusion image. The experimental verification results of the SAR images from the Sentinel-1 satellite and the multispectral images from the Landsat-8 satellite show that the proposed method can achieve superior visual effects, and excellent quantitative performance in terms of spectral distortion, correlation coefficient, MSE, NIQE, BRISQUE, and PIQE.
\end{abstract}

\begin{IEEEkeywords}
Synthetic aperture radar, multispectral image, remote sensing, pseudo-color fusion, coupled dictionary learning, Brovey transform
\end{IEEEkeywords}
\end{minipage}}}
\maketitle

\section{Introduction}
\label{sec:1}
\IEEEPARstart{W}{ith}
the expeditious developments of high-resolution Synthetic Aperture Radar (SAR) and multispectral remote sensing imaging equipment, earth observation technologies have improved remarkably. SAR imaging can achieve all-weather observations and detect the physical properties of the surface target (e.g., orientation, shape, roughness, the dielectric constant of the target, the frequency and incidence angle of the illuminating electromagnetic radiation)~\cite{chen2010sar}. However, SAR images acquired by remote sensors often suffer from geometric distortion, speckle noise, radio frequency interference, and other image degradation problems. Multispectral remote sensing can discriminate features based on the difference in morphology, structure of images, and spectral properties. It significantly expands the information volume of remote sensing~\cite{ghassemian2016review, ghamisi2019multisource}, and can be used for thematic mapping applications (e.g., land use surveys, soil erosion production). Researchers have paved the way for the pseudo-color fusion of SAR and multispectral in the past many years, and the fusion of these modalities can (i) provide a high spatial and high spectral resolution~\cite{ghamisi2019multisource}; (ii) assign each pixel to a specific class of interest, when making the map based on various modalities of remote sensing resources, resulting in better interpretation and more accurate, robust results in the fused image~\cite{piella2003general, lewis2004region}; and (iii) perform heterogeneous image fusion on massive remote sensing data to reduce the consumption of computing resources for deploying downstream tasks at the edge, and improve usability~\cite{ghassemian2016review}. The fused images will have complementary information and are used in various remote sensing applications (e.g., urban mapping, vegetation identification, and lithology analysis)~\cite{wald1999some}. This paper investigates improvements in the quality of SAR images assisted by multispectral imaging based on heterogeneous image fusion. The proposed method provides technical support for multi-source high-resolution earth observation~\cite{kulkarni2021application}.

Image fusion techniques include three different levels: pixel level, feature level, and decision level. This paper mainly focuses on pixel-level image fusion, which directly uses the primary information from the SAR and multispectral images~\cite{li2017pixel}.
Many algorithms for pixel-level image fusion~\cite{bovolo2009analysis,otazu2005introduction,nunez1999multiresolution,klonus2007image,chen2010sar,dalla2015global, ahmed2020sar, sun2014nearest,yilmaz2020genetic,singh2020response, jinju2019spatial} have been explored for remote sensing applications.  
Some approaches (e.g., principal component analysis (PCA), intensity-hue-saturation (IHS)) use SAR images to replace the optimal/derived band directly. Nevertheless, different bands shall have different wavelength coverage and operational principles, which will miss important information. Besides, some methods (e.g., Brovey) have an apparent tendency in the fusion target, and therefore cannot simultaneously retain the information in SAR and multispectral images.

As machine learning techniques develop rapidly~\cite{bai2022transformer,peleg2014statistical,yang2012coupled,wu2022two,veshki2020image,bai2021influence}, learning-based methodologies can be employed to explore the best intermediate representation of heterogeneous modalities~\cite{bai2023surgical,veshki2022multimodal,bai2023revisiting}.
Dictionary learning has achieved superior performance among various real-world applications~\cite{mairal2009online,kreutz2003dictionary,tovsic2011dictionary}. 
Sparse coefficients and over-complete dictionaries can be combined to reconstruct images through sparse representation. Yang et al. applied the theory to image fusion for the first time, and they tried to fuse multi-focal images using a discrete cosine transform dictionary-based method~\cite{yang2009multifocus}, as well as a simultaneous orthogonal matching pursuit-based method to fuse multimodal images~\cite{yang2012pixel}. Both Yin~\emph{et al.}~\cite{yin2011multimodal} and Yu~\emph{et al.}~\cite{yu2011image} attempted to represent source images as common and innovative components for image fusion. Kin~\emph{et al.} have combined joint patch clustering with dictionary learning for multimodal image fusion~\cite{kim2016joint}. As single dictionary models have been studied intensively, coupled dictionaries are required to represent dual feature spaces, such as two images with different resolutions or from heterogeneous sources. Coupled Dictionary Learning (CDL) has been applied to reconstruction~\cite{yang2012coupled}, recognition~\cite{ramirez2010classification, peleg2014statistical}, and signal fusion~\cite{veshki2020image}. Veshki~\emph{et al.}~\cite{veshki2022multimodal} proposed a CDL method based on simultaneous sparse approximation and relaxed the assumption of equal sparse representation. 
Zhang~\emph{et al.}~\cite{zhang2023joint} further employed CDL to preserve the structure, function, and edge information in the source images, overcoming the single dictionary's disadvantage. As a result of the rapid development of deep learning technologies and continuous advancements in high-performance computing, deep learning-based approaches have achieved excellent performance in remote sensing image fusion~\cite{luo2023dafcnn,jha2023gaf,deng2023psrt}. However, in order to achieve high performance, deep learning methods require training on large-scale datasets. Moreover, deep learning methods consume significant computational resources in terms of inference speed, storage space, and training costs, making them unsuitable for applications and deployment on edge devices such as satellites~\cite{zhang2020deep}. In comparison, CDL and traditional methods do not require large amount of training data, and the resulting coupled dictionaries occupy much less storage space. Additionally, sparse representation also enables high inference speed. Therefore, our paper will solely focus on discussing non-deep learning methods to cater to the demands of edge computing and deployment.

As a result of the rapid development of deep learning technologies and continuous advancements in high-performance computing~\cite{che2022learning,wu2023transformer,zia2023surgical,bai2023cat,che2023iqad}, deep learning-based approaches have achieved excellent performance in remote sensing image fusion~\cite{luo2023dafcnn,jha2023gaf,deng2023psrt}. However, in order to achieve high performance, deep learning methods require training on large-scale datasets. Besides, deep learning methods consume significant computational resources in terms of inference speed, storage space, and training costs, which restricts their usage to offline applications, and makes it challenging to deploy and operate them on edge devices~\cite{zhang2020deep}. In comparison, CDL and traditional methods do not require a large amount of training data, and the resulting coupled dictionaries occupy much less storage space. Additionally, sparse representation also enables high inference speed. Therefore, our paper will solely focus on discussing non-deep learning methods to cater to the demands of edge computing and deployment.

In the field of remote sensing applications, some studies have extended dictionary learning for image fusion, such as support value transformation and sparse representation~\cite{zhouping2015fusion}, and fusion-based cloud removal methods~\cite{huang2015cloud}. The CDL-based image fusion method can enforce learning to represent the relationship of two related feature spaces, aiming at two pairs of dictionaries to learn the same sparse representation. Therefore, CDL can spatially capture the dependency information of two images, and obtain a reasonable linear mathematical relationship between SAR and multispectral images, thus preserving different information from heterogeneous images (e.g., spatial information in SAR images and spectral information in multispectral images) and obtaining a better visual effect. Wang~\emph{et al.} further introduced the details injection model~\cite{wang2018pansharpening}. Ayas~\emph{et al.} considered introducing the texture information in the high-resolution image into the low-resolution image to enhance the effect of image fusion~\cite{ayas2018efficient}. CDL has also been used for collaborative prediction of multimodal remote sensing images, such as methods with distance preserved probability distribution adaptation~\cite{guo2022integrating}, and class-based guidance solutions~\cite{liu2022class}. 

This paper proposes a novel image fusion methodology with CDL and hybrid techniques to achieve the pseudo-color fusion of SAR images with multispectral images. Firstly, since the existing fusion solutions (selective mask, weighted fusion) will lead to information loss or attenuation when applied to multimodal data with obvious feature differences, we have applied the Brovey method to perform pre-processing and replace the original input SAR images. Thus, the ``pseudo'' SAR image shall contain information from two modalities, effectively avoiding information loss. Secondly, to strengthen the relationship between the two modalities in the coupled dictionary, we force the dictionaries to learn joint sparse representations. Meanwhile, we do not introduce restriction items when updating the dictionary, which can ensure the structural coherence of the coupled dictionaries and further promote the associativity of multimodal information. Lastly, the reconstruction error-based selection method is employed to generate the reconstruction mask for final fusion.
Our main contributions and findings are three-fold:

\begin{itemize}
    \item We introduce a novel hybrid algorithm, which integrates CDL and sparse representation to perform the pseudo-color fusion of SAR and multispectral images for the first time. This solution can capture the mutual relationship and establish the best intermediate representation, which efficiently generates fused images with rich spectral information, and geometric properties of SAR.
    \item We use Brovey transform as a pre-processing method and employ the Brovey image as the ``pseudo" SAR image with certain spectral information, enabling the final fused image with more comprehensive information.
    \item In the employed CDL algorithm, the coupled dictionaries are enforcedly learned together by the joint optimization scheme, and the structurally coherent dictionaries are also set by removing restriction items. Therefore, the multimodal correlation in the coupled dictionaries is further promoted, making it suitable for the multimodal fusion task on complex features.
    \item Experimental results on SAR images from the Sentinel-1 satellite and multispectral images taken by the Landsat-8 satellite demonstrate that our method can obtain remote sensing images with comprehensive spatial and spectral information, and achieve excellent fusion performance both qualitatively and quantitatively.
\end{itemize}

The rest of the paper is organized as follows: Section~\ref{sec:2} introduces the related work of SAR and multispectral image fusion; Section~\ref{sec:3} provides the formulation and details of the proposed methodology; the experiment results are presented in Section~\ref{sec:4} before drawing conclusions in Section~\ref{sec:5}.

\section{Related Works}
\label{sec:2}
Four categories may be used to categorize current pixel-level SAR and multispectral image fusion methodologies: component substitution, multiresolution analysis, hybrid techniques, and model-based algorithms~\cite{ghassemian2016review}. 

Component substitution methods include Brovey transform~\cite{earth1990brovey, bovolo2009analysis, vrabel1996multispectral}, Gram-Schmidt (GS)~\cite{laben2000process, dalla2015global, yilmaz2020genetic}, PCA~\cite{pohl1998multisensor}, and IHS~\cite{carper1990use, alparone2004fusion, alparone2004landsat}. Such methods will separate spatial components from spectral information in multispectral images, project them into another space, and use SAR images to replace the spatial components. The replaced image is then converted back to the original image space to acquire the fused result, which has high fidelity and renders spatial details in fused images.
However, the spectrum between the multispectral and SAR image channels is not matched, which causes local differences between the images. The component substitution method cannot account for this problem of local differences and can cause significant spectral distortion.

Multiresolution analysis methods need to decompose the original image into multiple scale levels based on wavelet transform~\cite{zhou1998wavelet,nunez1999multiresolution, nason1995stationary, shensa1992discrete, jinju2019spatial, bai2023llcaps} or pyramid transform~\cite{burt1987laplacian}, and then fuse the heterogeneous images at each level. Finally, the images from each level are recombined into a fused image. Yuan~\emph{et al.}~\cite{yuan2022multi} further introduced edge-preserving filters and weighted back-projection to improve information fidelity during multi-resolution fusion. A recent approach also considered decomposing the source image into structure and texture layers and performing fusion separately~\cite{chen2022infrared}. This method will increase the consumption of computing resources and computational complexity, but it can achieve better results and performance in the localization in both the spatial and frequency domains.

Hybrid image fusion techniques combine the benefits of component substitution approaches and multiresolution analysis approaches~\cite{chen2010sar,hong2009wavelet,luo2008fusion,yao2023rnn,kurban2022region,zhang2022multispectral,ghamisi2019multisource}. Chen~\emph{et al.}~\cite{chen2010sar} used trous wavelet transform decomposition (AWT) to extract the details of SAR images and applied empirical mode decomposition (EMD) to discern high-frequency information in multispectral images and the approximate images of SAR. Finally, an additive operation was conducted in the AWT-EMD domain to achieve the final fused high-resolution image. Hong~\emph{et al.}~\cite{hong2009wavelet} combined wavelet and IHS fusion to maximize the color and spatial information from source images. Luo~\emph{et al.}~\cite{luo2008fusion} integrated PCA and additive wavelet decomposition, attempting to solve the decreased spatial resolution in wavelet fusion shall sacrifice and the severe spectral distortion in PCA fusion. Kurban~\cite{kurban2022region} analyzed the image fusion problem as an optimization problem, combining differential search theory with IHS transform. Zhang~\emph{et al.}~\cite{zhang2022multispectral} connected the Laplacian pyramid with sparse representation. They decomposed the source images into high-frequency and low-frequency components, and fused them by sparse representation, respectively. Combining the above two methods, hybrid techniques can generate fused images with different characteristics and emphasis towards different remote sensing applications~\cite{ghamisi2019multisource}.

Model-based algorithms present excellent capabilities for representing the complicated local features of remote sensing data. A variety of model-based methods have been applied to SAR and multispectral image fusion, such as Bayesian estimation~\cite{wei2015bayesian}, compressive sensing~\cite{ghahremani2015compressed}, CDL~\cite{guo2014online}, sparse representation~\cite{wei2015hyperspectral}, and deep neural network (DNN)~\cite{li2017pan}. Camacho~\emph{et al.}~\cite{camacho2022hyperspectral} employed an augmented linear mixing model to deal with the spectral variability problem in image fusion. Sun~\emph{et al.}~\cite{sun2022feature} introduced the ant colony optimization algorithm to explore the global optimal path in image fusion. 
Sparse representation solutions have also been widely explored in multimodal image fusion (e.g., near infrared-visible image pairs, visible-infrared image pairs, PET-MRI image pairs, and multi-focus image pairs)~\cite{wang2017novel,zhu2018novel}. Wang~\emph{et al.}~\cite{wang2017novel} employed geometric information to train image patches as sub-dictionaries and combined input image pairs using the proposed constructed-sub-dictionary (CSD) strategy. Zhu~\emph{et al.}~\cite{zhu2018novel} trained dictionaries using clustering algorithms, fused images using sparse representation to preserve texture details, and employed energy-based spatial algorithms to retain structural information during the fusion process. However, a more effective approach is to directly create coupled dictionaries during the process of dictionary learning. This allows for maximum preservation of the respective valuable information from different modalities, while effectively capturing the correlation between the two modalities.
Model-based methods allocate pixels of the fused images based on different fitting strategies, seek the optimal combination of spectral and spatial information, and obtain competitive fusion results.

In this paper, the proposed pseudo-color fusion technique combines the advantages of component substitution methods, hybrid techniques, and model-based methods. We initially use the Brovey method to pre-process multispectral and SAR images, to endue the ``pseudo'' SAR image with spectral information. Different from traditional combination approaches of hybrid techniques, CDL is employed to integrate Brovey and multispectral images. Model-based methods have the capability of exploring the best intermediate representation of input data. Hybrid techniques can summarize the advantages of both component substitution and multiresolution analysis methods, while our method can aggregate benefits from these three different types of methods. Our proposed novel method is expected to effectively combine and complement the information in SAR and multispectral images, and to produce high-quality fused images.

\section{Methodology}
\label{sec:3}

\begin{figure*}[t]
    \centering
    \includegraphics[width=1\textwidth, trim=0 0 240 0]{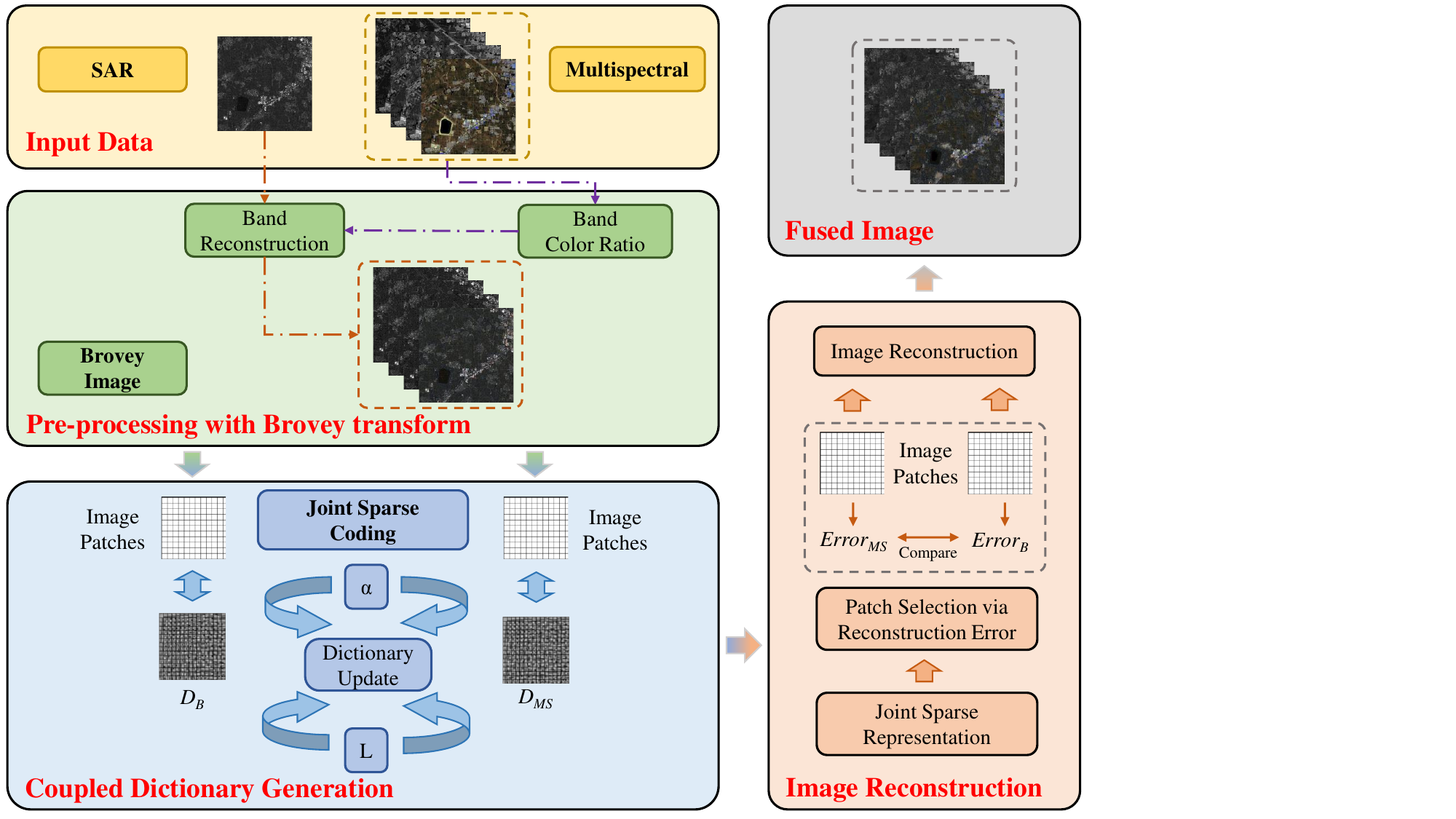}
    \caption{
    Flowchart of the proposed pseudo-color fusion methodology. Pre-processing with Brovey transform endues the ``pseudo" SAR image with certain spectral information. CDL is employed to capture the mutual relationship and establish the coupled dictionaries of the multispectral and Brovey images. Finally, the image is reconstructed via joint sparse representation.
    }
    \label{fig:1}
\end{figure*}

This work first pre-processes the input SAR image $I_{SAR} \in R^{b \times c}$ and multispectral image $I_{MS} \in R^{b \times c}$ using the Brovey fusion method, and get the Brovey image  $I_B \in R^{b \times c}$. Then, CDL is used to analyze the correlation properties of Brovey and multispectral images. Finally, the reconstruction errors of a pair of input patches are set as the discriminative rule for image patch reconstruction. Briefly, for the Brovey images and multispectral images, we extract $\alpha$ patches of size $p$ from these two images as matrices $X_B \in R^{p \times q}$, and $X_{MS} \in R^{p \times q}$, respectively. Then we select the appropriate patches from the image fused via Brovey transform to replace the corresponding patches in the multispectral matrix, and stitch to get a final matrix $X_{Fusion} \in R^{p \times q}$. A weighted average process is devised for the edges of the stitched patches. The workflow of the algorithm is shown in Fig.~\ref{fig:1}, and the pseudo-code is presented in Algorithm~\ref{alg:1}.

\subsection{Pre-processing via Brovey Transform}

The Brovey fusion algorithm~\cite{earth1990brovey} is a type of ratio-transform fusion, which normalizes each band of a multispectral image, and then performs a multiplicative band operation on a panchromatic image. As the Brovey transform operation is related to the color space, this algorithm can only be conducted on remote sensing images with three bands~\cite{vrabel1996multispectral, bovolo2009analysis}. This methodology was originally designed for panchromatic and multispectral fusion. In our pseudo-color fusion application, we employ SAR images as pseudo-panchromatic images.
Brovey transform can decompose the image elements of a multispectral image into color and luminance separately. The expression of the Brovey fusion algorithm is shown as follows:
\begin{equation}
\begin{split}
\begin{aligned}
    & {I_B}_{i} = {I_{MS}}_{i} \times {I_{SAR}} / \theta \\
    & \theta= \sum_{\beta=1}^n {I_{MS}}_{i}
\end{aligned}
\end{split}
\label{equ:1}
\end{equation}
where $I_{MS}, I_{SAR}, \theta \in R^{b \times c}$. $i$ represents the three bands of the multispectral and Brovey images, and therefore $n$ = $3$. ${I_B} \in R^{b \times c}$ denotes the fused Brovey image. Therefore, based on the principle of chromaticity transform~\cite{roberts2008assessment}, Brovey transform possesses the ability to preserve a high degree of spatial detail through arithmetical technique with the SAR image. The defect of the Brovey transform is that the results of Brovey fusion usually contain high spectral distortion and lack spectral information. However, in other words, the image pre-processed with Brovey transform is highly similar to the SAR image, while also containing certain spectral information. From the visual evaluation, the Brovey transform imparts ``color" information to the SAR image. In this case, we utilize this ``defect" and apply the Brovey image as the ``pseudo" SAR image in the subsequent fusion step. Then, we can further explore the potential of the Brovey transform using CDL and joint sparse representation.

\subsection{Coupled Dictionary Generation}
After achieving the multispectral image patches $X_{MS}$ and another pre-processed image patches $X_B$ using the Brovey method, we generate a pair of coupled dictionaries $D_{MS}$, $D_B$, which represent the set of the input data $X_{MS}$, $X_B$, respectively. The coupled dictionary generation procedure can be formulated as follows:
\begin{equation}
\begin{split}
    \min _{D_{MS}, D_B, L}\left\|X_{MS}-D_{MS} L\right\|_F^2+\left\|X_B-D_B L\right\|_F^2 \\
    s.t. \left\|\alpha_m\right\|_0 \leq H_0,\left\|\left[d_{MS}\right]_n\right\|_2=1,\left\|\left[d_B\right]_n\right\|_2=1, \forall n, m
\end{split}
\label{equ:cdg1}
\end{equation}
where $[d_{MS}]_n$ and $[d_B]_n$ refer to the $n^{th}$ columns of the atoms. $\| \cdot \|_0$ represents the number of non-zero coefficients. $\| \cdot \|_2$ is the Euclidean norm, and $\| \cdot \|_F$ is the Frobenius norm. $H_0$ is a constraint coefficient. $L$ denotes the common sparse representation matrix. Therefore, the two dictionaries from different modalities are enforced to be learned together under one optimization procedure, which promotes their learning on joint sparsity representation. Equ.~(\ref{equ:cdg1}) can then be rewrote as:
\begin{equation}
    \min _{D_{MS}, D_B}\left\|X_{MS}-\sum_{n}[d_{MS}]\left[\alpha_n^T\right]\right\|_F^2+\left\|X_{B}-\sum_{n}[d_{B}]\left[\alpha_n^T\right]\right\|_F^2
\label{equ:cdg2}
\end{equation}
The relationship between the multispectral and Brovey image can thereby be captured for further updating and reconstruction.

We use the coupled dictionary training method proposed in~\cite{veshki2019efficient} based on an iterative minimization approach to solve~(\ref{equ:cdg2}) and thus obtain the pair of dictionaries $D_{MS}$ and $D_B$. The method minimizes the dictionary pair and the sparse coding $L$ alternatively. Sparse encoding is performed using orthogonal matching pursuit (OMP)~\cite{tropp2007signal}. The dictionary is optimized by minimizing the below item:
\begin{equation}
\left[d_r\right]_n=\underset{\left[d_r\right]_n}{\operatorname{argmin}}\left\|\left[E_r\right]_n-\left[d_r\right]_n\left[\alpha_n^T\right]_{\int_n}\right\|_F^2, r \in\{{MS}, B\}
\label{equ:3}
\end{equation}
\begin{equation}
    \left[E_r\right]_n \triangleq\left[X_r-\sum_{t \neq n}\left[d_r\right]_t \alpha_n^T\right]_{\alpha_n}, \; \int_n=\left\{m \mid \left[\alpha_n^T\right]_m \neq 0\right\}
\label{equ:4}
\end{equation}
in which $\alpha_n^T$ denote the $n^{th}$ row of $L$, and $\int_n$ is used to select the non-zero entries in $\alpha_n^T$. The subscript ${MS}$ and $B$ denote the multispectral image and the image pre-processed with the Brovey method, respectively. Reference~\cite{veshki2020image} imposes constraints on dictionary updates such that the pair of coupled dictionaries can be discriminated. However, in our remote sensing application, we employ the CDL setup from~\cite{veshki2019efficient}, using two separate least squares and $\mathcal{L}_2$-norm to update the atom. Meanwhile, we remove the restriction items in~\cite{veshki2020image} to further enhance the correlation between the two dictionaries.
When $\int_n = \emptyset$, the updated value of $\left[d_r\right]_n$ is obtained by computing the column-wise average of error matrix $\left[E_r\right]_n$:
\begin{equation}
\left[E_r\right]_n = X_{r}-D_{r} L, \; r \in\{{MS}, B\}
\label{equ:empty}
\end{equation}
When $\int_n \neq \emptyset$, the $\mathcal{L}_2$-norm is then employed for updating and optimization:
\begin{equation}
\left[d_r\right]_n= \frac{\left[E_r\right]_n\left[\alpha_n^T\right]^T_{\int_n}}{\|\left[\alpha^T_n\right]_{\int_n}\|^2_2}, \; r \in\{{MS}, B\}
\label{equ:atom1}
\end{equation}
We employ the $\mathcal{L}_2$-norm here, so we need to normalize the normalization term $\|\left[\alpha^T_n\right]_{\int_n}\|^2_2$ to $1$, so the Equ.~(\ref{equ:atom1}) can be rewrote as:
\begin{equation}
\left[d_r\right]_n=\left[E_r\right]_n\left[\alpha_n^T\right]^T_{\int_n}, \; r \in\{{MS}, B\}
\label{equ:atom2}
\end{equation}
Subsequent to the updating of $\left[d_r\right]_n$, based on the intial setup in Equ.~(\ref{equ:cdg2}), the $\left[\alpha^T_n\right]_{\int_n}$ can also be updated as:
\begin{equation}
\left[\alpha^T_n\right]_{\int_n} = {d_n^T}{E_n}
\label{equ:atom3}
\end{equation}
\begin{equation}
{d_n} \triangleq {[d_{MS}]_n}/{[d_B]_n},  \; {E_n} \triangleq {[E_{MS}]_n}/{[E_B]_n}
\label{equ:atom4}
\end{equation}
Therefore, we construct the coupled dictionaries of double feature spaces from the multispectral and Brovey image pairs.

\begin{algorithm}[t] 
	\caption{Multispectral and SAR Image Pseudo-color Fusion} 
	\label{alg:1}
		{\bf Require:} Input multispectral image $I_{MS}$ and SAR image $I_{SAR}$, $D_{MS}=D_{B}=D_0$. \\
	01: \hspace*{0.1in} Obtain $I_B$ via Brovey transform-based image fusion with Equ.~(\ref{equ:1}); \\
        02: \hspace*{0.1in} Extract patch matrices $X_{MS}$ and $X_B$; \\
        03: \hspace*{0.1in} Normalize patches $x_{MS}$ and $x_B$; \\
        04: \hspace*{0.1in} {\bf for} {R rounds of updating} {\bf do}: \\
        05: \hspace*{0.1in} \hspace*{0.1in} Compute $L$ in the sparse coding with Equ.~(\ref{equ:cdg1}); \\
        06: \hspace*{0.1in} \hspace*{0.1in} {\bf for} {number of atoms $n=1,2,...$} {\bf do}: \\
        07: \hspace*{0.1in} \hspace*{0.1in} \hspace*{0.1in} Compute $\int_n$ with Equ.~(\ref{equ:4}); \\
        08: \hspace*{0.1in} \hspace*{0.1in} \hspace*{0.1in} {\bf if}{$\int_n = \emptyset$} {\bf then}: \\
        09: \hspace*{0.1in} \hspace*{0.1in} \hspace*{0.1in} \hspace*{0.1in} Update $[d_r]_n$ with Equ.~(\ref{equ:empty}); \\
        10: \hspace*{0.1in} \hspace*{0.1in} \hspace*{0.1in} {\bf else}: \\
        11: \hspace*{0.1in} \hspace*{0.1in} \hspace*{0.1in} \hspace*{0.1in} Update $[d_r]_n$ with Equ.~(\ref{equ:atom2}); \\
        12: \hspace*{0.1in} \hspace*{0.1in} \hspace*{0.1in} \hspace*{0.1in} Update $\left[\alpha^T_n\right]_{\int_n}$ with Equ.~(\ref{equ:atom3}); \\
        13: \hspace*{0.1in} \hspace*{0.1in} \hspace*{0.1in} {\bf end if} \\
        14: \hspace*{0.1in} \hspace*{0.1in} {\bf end for} \\
        15: \hspace*{0.1in} {\bf end for} \\
        16: \hspace*{0.1in} Obtain the dictionaries $D_{MS}$ and $D_{B}$; \\
        17: \hspace*{0.1in} Establish the coupled dictionary $D$ with Equ.~(\ref{equ:6}); \\
        18: \hspace*{0.1in} {\bf for} ($x_{MS}$ and $x_B$) in ($X_{MS}$ and $X_B$) {\bf do}: \\
	19: \hspace*{0.1in} \hspace*{0.1in} Calculate $e_{MS}$ and $e_B$ to establish $K_\alpha$ with Equ.~(\ref{equ:7}); \\
	20: \hspace*{0.1in} {\bf end for} \\
        21: \hspace*{0.1in} Compute the reconstruction mask $K$ with Equ.~(\ref{equ:8}); \\
        22: \hspace*{0.1in} Fuse $I_{MS}$ and $I_B$ and obtain $I_F$ with Equ.~(\ref{equ:9}).
\end{algorithm}

\subsection{Pseudo-color Fusion via Sparse Representation}
Subsequent to constructing the coupled dictionaries of multispectral image and Brovey image, we then classify the image signal with sparse representation. Through evaluating the reconstruction errors, each element is able to be assigned with the minimum reconstruction errors to obtain the best sparse representation classification~\cite{skretting2006texture, veshki2020image}. Firstly, a dictionary $D$ is built based on the concatenation of the obtained pair of coupled dictionaries:
\begin{equation}
    D \triangleq\left[\begin{array}{ll}
D_{\mathrm{MS}}^T & D_{\mathrm{B}}^T
\end{array}\right]^T
\label{equ:6}
\end{equation}

Then, we use $e_{MS}$, $e_B$ to denote the reconstruction errors of the multispectral images and the pre-processed images using the Brovey method. The reconstruction error formulation is as follows:
\begin{equation}
   \left\{\begin{array}{l}
    e_{MS}=\left\|D \alpha-\left[x_B x_{MS}\right]\right\|_2^2 \\
    e_B=\left\|D \alpha-\left[x_{MS} x_B\right]\right\|_2^2
    \end{array}\right. 
\label{equ:7}
\end{equation}
in which $\alpha$ includes the related sparse codes. When $e_{MS} < e_B$, $x_{MS}$ will be chosen, otherwise $x_B$ will be chosen. $e_{MS}$ and $e_B$ will not be equal~\cite{veshki2020image}, so there is no need to consider the case when they are equal. By applying this selection rule to all patches, the patch-level mask $K_\alpha$ can then be obtained directly. The pixel-level mask $K$ is able to be calculated with $K_\alpha$ by applying the following equation:
\begin{equation}
    K=P^*\left(K_\alpha\right)-1
\label{equ:8}
\end{equation}

$K_\alpha$ contains values within an interval $\{1, 2\}$, so the values in $K$ are within $[0,1]$. $P^{*}\left( \cdot \right)$
is a function that can place each patch in the image where it should be located. This function performs weighted averaging operations for overlaps between patches. Therefore, we can achieve the final fused image using the above mask as follows:
\begin{equation}
    I_{F, u, v}=K_{u, v} I_{{MS}, u, v}+\left(1-K_{u, v}\right) I_{B, u, v}
\label{equ:9}
\end{equation}
$I_F$ denotes the finally fused image, and $(u,v)$ denotes the location of pixels in the image.

\subsection{Complexity}
Firstly, the steps of Brovey fusion, mask generation, and final fusion take up little computational cost, as they are only a small amount of pixel-level or patch-level element-wise addition and multiplication operations, far less than those from coupled dictionary generation. In this case, the coupled dictionary generation phase shall dominate the complexity calculation. Based on~\cite{song2019coupled}, the complexity of the sparse coding phase  is $O(pq{S_p}A{H_0}N)$ (in Equ.~(\ref{equ:3}) \& (\ref{equ:4})). $S_p$ denotes the patch size, $A$ denotes the number of atoms, and $N$ denotes the number of iterations during the optimization. Besides, the complexity during the atom updating period is $2 \times O(pq)$ (in Equ.~(\ref{equ:atom2}) \& (\ref{equ:atom3})). Therefore, the total complexity during the coupled dictionary generation is the sum of the sparse coding and atom updating phase.

\section{Experiments}
\label{sec:4}
\subsection{Implementation Details}
All experiments are conducted on a computer with AMD Ryzen™ 9 5950X 3.40GHz CPU and NVIDIA RTX 3090 GPU. 
Our SAR data are products from the Sentinel-1, which have two satellites to satisfy the requirements of revisit and coverage. 
The satellite usually scans in Interferometric Wide (IW) swath mode when observing the land. Data from this Sentinel-1 satellite is publicly available\footnote{\href{https://sentinel.esa.int/web/sentinel/toolboxes/sentinel-1}{sentinel.esa.int/web/sentinel/toolboxes/sentinel-1}}. Our multispectral data comes from the Operational Land Imager (OLI) on the Landsat-8 satellite, which is also publicly accessible\footnote{\href{https://glovis.usgs.gov/}{https://glovis.usgs.gov/}}. The Landsat-8 satellite collects images by capturing electromagnetic waves reflected and radiated from the earth's surface and converting them into digital signals. The data we used was taken in 2022 in Liangshan, Shandong Province, China. The two modality images are registered directly based on standard map coordinates in ENVI~5.5.3. The experimental results are compared qualitatively and quantitatively against wavelet-based fusion~\cite{jinju2019spatial}, PCA-based fusion~\cite{roberts2008assessment}, Hue-saturation-value (HSV) color space-based fusion~\cite{ahmed2020sar}, GS fusion~\cite{yilmaz2020genetic}, Nearest-Neighbor Diffusion (NND)-based Fusion~\cite{singh2020response}, improved wavelet transform (Improved Wavelet)-based fusion~\cite{richa2022effective}, adaptive enhanced fusion (AEF)~\cite{yin2022adaptive}, and coupled feature learning via structured convolutional sparse coding (CCFL)~\cite{veshki2022coupled}.
While maintaining the integrity of the classical methods, we select the latest methods of the same type for comparison. These methods can be referred to the corresponding references for details. As we discussed in Section II, to ensure the performance of algorithms computing and deployment at the edge, we did not introduce deep neural network algorithms that require high training consumption.

\subsection{Evaluation Metrics}

To evaluate the image quality, we adopt reference-based metrics, including the degree of spectral distortion~\cite{chen2010sar}, correlation coefficient~\cite{xiong2018order}, mean-square error (MSE)~\cite{ahmed2022pseudo}, and no-reference metrics including Natural Image Quality Evaluator (NIQE)~\cite{mittal2012niqe}, Blind/Referenceless Image Spatial QUality Evaluator (BRISQUE)~\cite{mittal2012brisque}, Perception-based Image Quality Evaluator (PIQE)~\cite{venkatanath2015piqe}. We shall present the equations of reference-based metrics, and the detailed algorithms of the no-reference metrics can be referred to in~\cite{mittal2012niqe, mittal2012brisque, venkatanath2015piqe}, respectively.

\subsubsection{Degree of Spectral Distortion}
When comparing with the original multispectral image, the obtained image's level of distortion is directly represented by the degree of spectral distortion~\cite{li2019multi, chen2010sar}.
\begin{equation}
    \boldsymbol{\rm Distortion}=\frac{1}{N} \sum_{i=1}^N\left|Fuse(i)-Ref(i)\right|
\end{equation}
where $i$ stands for each pixel in the image and $N$ denotes the number of pixels. $Ref(i)$ and $Fuse(i)$ represent the pixels of the reference and fused images. The multispectral image is set to the reference. The higher spectral distortion represents the more severe spectral distortion.

\subsubsection{Correlation Coefficient}
The correlation coefficient, within the range of $[-1,1]$, measures the similarity between reference and fused images~\cite{xiong2018order}. It is evaluated as follows:
\begin{equation}
\boldsymbol{\rm CC}=\frac{\sum_{i=1}^N [Ref(i)-\overline{Ref}] \cdot[Fuse(i)-\overline{Fuse}]}{\sqrt{\sum_{i=1}^N [Ref(i)-\overline{Ref}]^2 \cdot[Fuse(i)-\overline{Fuse}]^2}}
\end{equation}
where $\overline{Ref}$ and $\overline{Fuse}$ represent the mean value of the pixels.
The higher the $\boldsymbol{\rm CC}$, the more correlated the two images are.

\subsubsection{MSE}
MSE calculates the mean square root error of the three bands between the fused and reference images, which measures the similarity rate~\cite{ahmed2022pseudo}.
\begin{equation}
    \boldsymbol{\rm MSE} = {\frac{1}{N} \sum_{i=1}^N \left[Fuse(i)-Ref(i)\right]^2}
\end{equation}

We also employ the multispectral image as the reference.

\begin{figure*}[t]
    \centering
    \includegraphics[width=1\textwidth, trim=0 30 345 0]{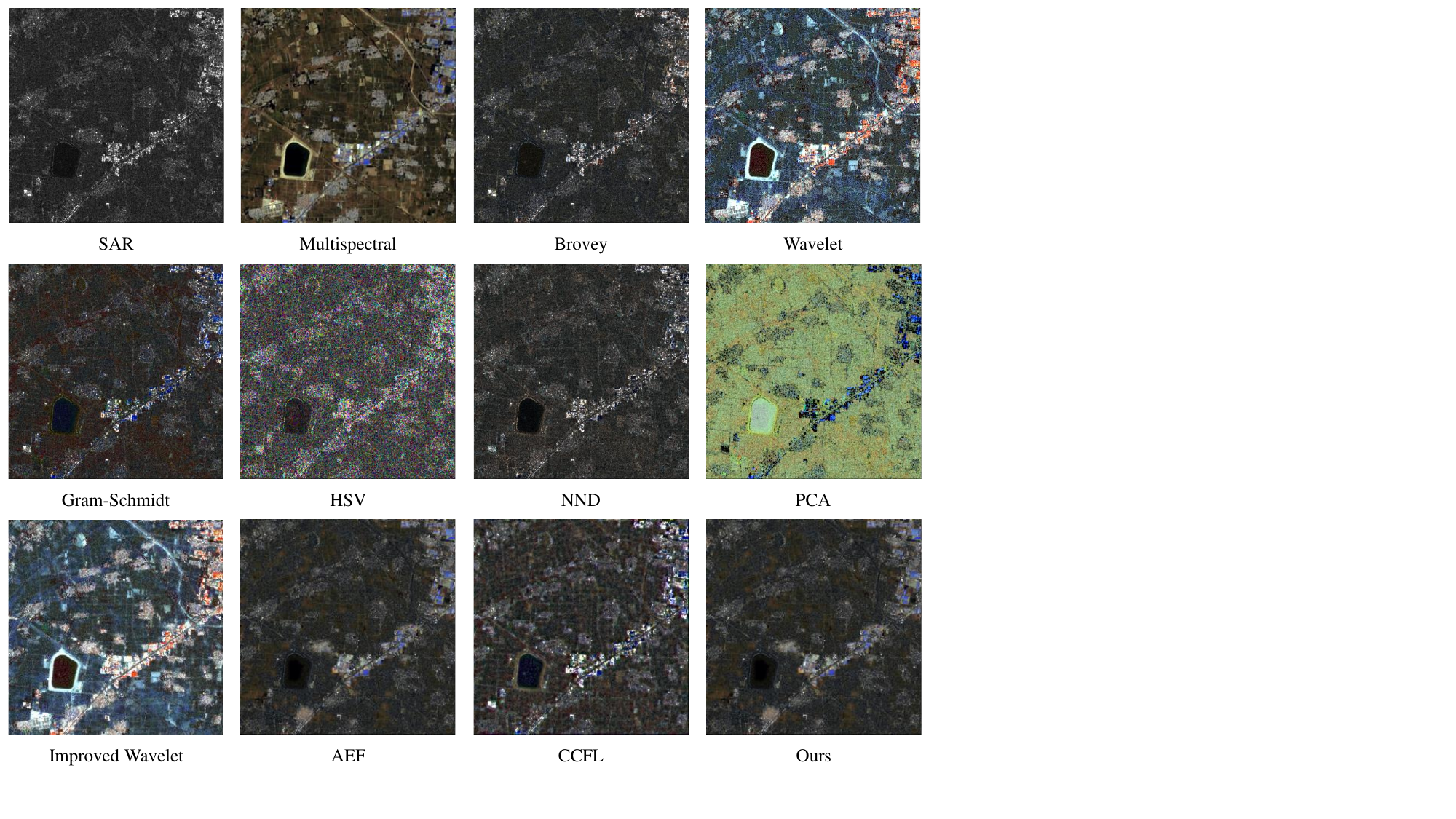}
    \caption{Qualitative comparison experiment demonstration of our pseudo-color fusion methodology, with the methods based on Wavelet, GS, HSV, NND, PCA, Improved Wavelet, AEF, CCFL together with the original SAR image, multispectral image, and image fused with Brovey method.
    }
\label{fig:demo1}
\end{figure*}   

\begin{figure*}[t]
    \centering
    \includegraphics[width=1\textwidth, trim=0 30 345 0]{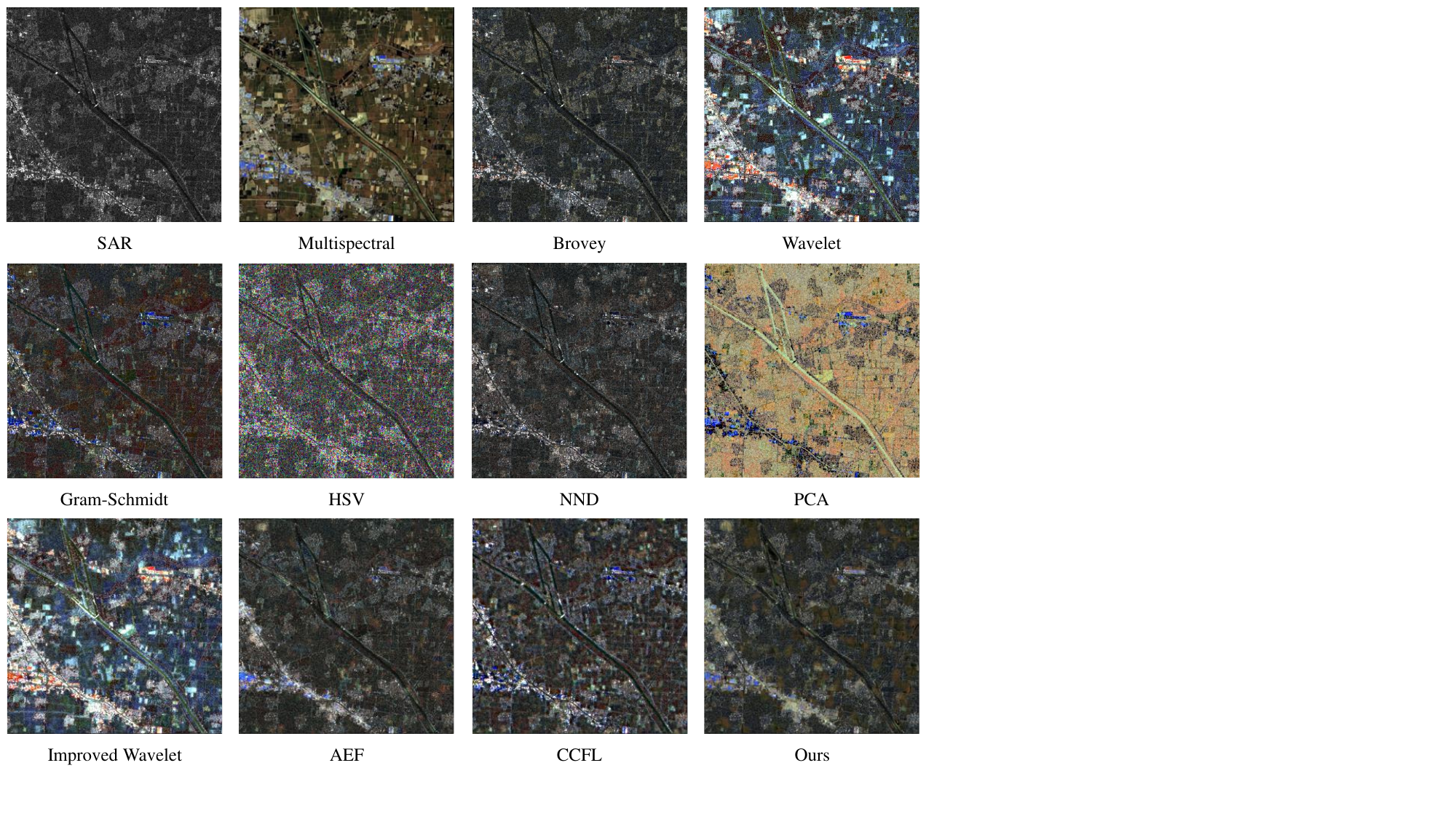}
    \caption{Qualitative comparison experiment demonstration of our pseudo-color fusion methodology, with the methods based on Wavelet, GS, HSV, NND, PCA, Improved Wavelet, AEF, CCFL together with the original SAR image, multispectral image, and image fused with Brovey method.
    }
\label{fig:demo2}
\end{figure*}

\subsection{Experimental Results}

The performance of our proposed fusion methodology is both quantitatively (Table~\ref{tab:1} and \ref{tab:2}) and qualitatively (Fig.~\ref{fig:demo1} and~\ref{fig:demo2}) benchmarked against the commonly used remote sensing image fusion methods of Wavelet, GS, HSV, NND, PCA, Improved Wavelet, AEF, and CCFL.

Firstly, we discuss the subjective visual evaluation of all the fusion techniques. As we discussed, the Brovey images are highly similar to the SAR images and contain a small amount of spectral information (namely, color information in the RGB color space). The HSV images introduce severe scattered noise, and the PCA images cause serious color distortion from the perspective of the RGB color space. Although the images fused using the wavelet method and the improved wavelet method show significant differences from the original SAR and multispectral images, the wavelet images still maintain a high degree of recognition in terms of color, texture, resolution, etc., and we will further discuss the results of the wavelet methods in the consecutive quantitative analysis. GS, NND, and our proposed approach have better fidelity to the original heterologous information. Nevertheless, the speckle white noises from the SAR images remain in the GS and NND images. Our method removes these noise points effectively, making the resulting images smooth, highly observable, and free of noise pollution. The CDL-based methods (AEF and CCFL) present similar results to our method, but present some blurring and distortion in the brighter regions. In contrast, our method does well in brighter regions. In summary, our proposed pseudo-color fusion method achieves superior performance from the above visual evaluation.

Subsequently, we discuss the quantitative analysis of the above fusion methods. Regarding the degree of spectral distortion, our proposed fusion technique performs the best compared to all comparison approaches, proving that it possesses the most spectral fidelity.
Like the visual evaluation, PCA and HSV methods bring significant spectral distortions, while NND and GS present less spectral fidelity than ours. It is worth mentioning that the wavelet-based fusion approaches also have a good performance in spectral fidelity. AEF and CCFL also perform well in spectral fidelity, slightly lower than our solution.

The correlation coefficient evaluates the similarity between the fused and source images. We measure the correlation coefficient with multispectral and SAR images, respectively. The average correlation coefficients are also obtained to measure which fusion approach retains the most information from the two heterologous images. HSV and PCA still suffer from noises and distortion and perform poorly in the comparison study. Despite PCA achieving a relatively high correlation with SAR images, HSV and PCA perform poorly in all other metrics. The wavelet and improved wavelet methods maintain a high correlation with multispectral images and a low correlation with SAR images. Thus, it should not be considered the excellent intermediate representation of different modalities in our scene. GS, NND, AEF, CCFL, and our method fall into the final round. These three methods achieve excellent and similar performance in retaining information from the heterologous source. Regarding the correlation coefficient, NND shows the best performance, while our proposed methods achieve the second. Although our method does not achieve the best performance on any single band, our overall correlation coefficient ranks second, proving that our method can effectively aggregate multi-source information. The comparison in terms of MSE is straightforward, and our proposed pseudo-color fusion method achieves the best reconstruction MSE, followed by CCFL, NND, GS, wavelet, improved wavelet, and AEF, in order. HSV and PCA still have the worst performance in MSE.

Finally, our method achieves remarkable performance among all no-reference image quality assessment approaches, only second to CCFL in PIQE. In the performance of NIQE and BRISQUE, our method is always in the first place. The excellent values of NIQE and BRISQUE prove that our image has exceptional performance under the reference of a statistical model based on a large external natural scene corpus, indicating that the features of the image fused by our method are closer to the feature benchmark established by NSS. The high performance on PIQE proves that our solution can effectively prevent distortion between the local patches.

\begin{table*}[t]
    \renewcommand\arraystretch{0.4}
	\caption{
        Comparison experiments on the degree of spectral distortion, correlation coefficient \& MSE, against Wavelet, GS, HSV, NND, PCA, Improved Wavelet, AEF, and CCFL approaches. \textbf{Bold} and \underline{underline} represent the best and the second-best performance, respectively. In respect of the degree of spectral distortion and MSE, ``Overall" denotes the average results of three bands. In respect of the correlation coefficient, ``MS" denotes the average results of three multispectral bands, and ``Overall" denotes the average results of ``MS" and ``SAR".
	}
	\centering
	\label{tab:1}  
	\resizebox{\textwidth}{!}{
        \setlength{\tabcolsep}{1.5mm}{
	\begin{tabular}{c|cccc|cccccc|cccc}
		\noalign{\smallskip}\hline\noalign{\smallskip}	
        \multirow{2}{*}{\makecell[c]{Method}} & \multicolumn{4}{c|}{Degree of Spectral Distortion $\downarrow$} & \multicolumn{6}{c|}{Correlation Coefficient $\uparrow$} & \multicolumn{4}{c}{MSE $\downarrow$} \\
        \noalign{\smallskip}
        & Band 1 & Band 2 & Band 3 & Overall & Band 1 & Band 2 & Band 3 & MS & SAR & Overall & Band 1 & Band 2 & Band 3 & Overall \\
        \noalign{\smallskip}\hline\noalign{\smallskip}
        Wavelet & \textbf{25.58} & 25.29 & 44.68 & 31.85 
        & \textbf{0.7227} & \underline{0.8718} & \textbf{0.7393} & \textbf{0.7779} & 0.3502 & 0.5640 
        & \underline{107.40} & \textbf{99.41} & 159.98 & 140.31 \\
        \noalign{\smallskip}
        GS & 31.86 & 31.15 & 25.12 & 29.38 
        & 0.3272 & 0.3020 & 0.3583 & 0.3292 & \textbf{0.9599} & 0.6445 
        & 127.82 & 126.67 & 103.48 & 121.84 \\
	\noalign{\smallskip}
        HSV & 41.27 & 40.97 & 44.90 & 42.38
        & 0.1520 & 0.1876 & 0.2768 & 0.2055 & 0.6135 & 0.4095 
        & 159.90 & 159.01 & 168.72 & 178.22 \\
        \noalign{\smallskip}
        NND & 29.75 & \underline{27.83} & \underline{24.03} & \underline{27.20} 
        & 0.4910 & 0.4122 & 0.4387 & 0.4473 & \underline{0.9387} & \textbf{0.6930} 
        & 117.77 & 115.83 & 103.34 & \underline{115.80} \\
        \noalign{\smallskip}
        PCA & 81.20 & 84.33 & 62.51 & 76.02 
        & 0.1283 & 0.2219 & 0.1542 & 0.1681 & 0.8484 & 0.5083 
        & 270.81 & 281.47 & 211.74 & 256.96 \\
        \noalign{\smallskip}
        Improved Wavelet & 30.33 & 29.64 & 39.78 & 33.25 & \underline{0.6590} & \textbf{0.8827} & \underline{0.7377} & \underline{0.7398} & 0.5327 & 0.6363 & \textbf{102.48} & 113.66 & 151.06 & 122.40  \\\noalign{\smallskip}
        AEF & 31.70 & 29.97 & 24.65 & 28.77 & 0.3890 & 0.3570 & 0.4997 & 0.4152 & 0.8642 & 0.6397 & 150.32 & 128.50 & 100.48 & 126.43 \\\noalign{\smallskip}
        CCFL & \underline{29.44} & 28.98 & 27.11 & 28.52 & 0.3775 & 0.3225 & 0.4306 & 0.3769 & 0.9076 & 0.6422 & 132.70 & 115.64 & \underline{90.65} & \underline{113.00} \\
        \noalign{\smallskip}
        Ours & 29.75 & \textbf{25.10} & \textbf{18.20} & \textbf{24.35} 
        & 0.3598 & 0.4597 & 0.6725 & 0.4973 & 0.7940 & \underline{0.6456} 
        & 126.50 & \underline{111.66} & \textbf{83.14} & \textbf{106.03}\\
	\noalign{\smallskip}\hline
	\end{tabular}}}
\end{table*}

\begin{table*}[t]
    \renewcommand\arraystretch{0.4}
	\caption{
        Comparison experiments on NIQE, BRISQUE \& PIQE, against Wavelet, GS, HSV, NND, PCA, Improved Wavelet, AEF, and CCFL  approaches. \textbf{Bold} and \underline{underline} represent the best and the second-best performance, respectively.  ``Overall" denotes the average results of three bands.
	}
	\centering
	\label{tab:2}  
        \resizebox{\textwidth}{!}{	
        \setlength{\tabcolsep}{1.5mm}{
	\begin{tabular}{c|cccc|cccc|cccc}
		\noalign{\smallskip}\hline\noalign{\smallskip}	
        \multirow{2}{*}{\makecell[c]{Method}} & \multicolumn{4}{c|}{NIQE $\downarrow$} & \multicolumn{4}{c|}{BRISQUE $\downarrow$} & \multicolumn{4}{c}{PIQE $\downarrow$} \\
        \noalign{\smallskip}
        & Band 1 & Band 2 & Band 3 & Overall & Band 1 & Band 2 & Band 3 & Overall & Band 1 & Band 2 & Band 3 & Overall \\
        \noalign{\smallskip}\hline\noalign{\smallskip}
        Wavelet & 7.03 & 6.60 & 6.19 & 6.61 
        & 30.50 & 24.85 & 24.87 & 26.74	
        & 39.10 & 40.57 & 36.99 & 38.89 \\
        \noalign{\smallskip}
        GS & 5.42 & \underline{5.19} & \textbf{4.22} & \underline{4.94}
        & 24.85 & \underline{22.11} & \textbf{18.66} & \underline{21.87}
        & 39.51 & 38.83 & \underline{27.09} & \underline{35.14} \\
	\noalign{\smallskip}
        HSV & 9.41 & 9.26 & 10.01 & 9.56
        & 44.93 & 45.26 & 44.90 & 45.03
        & 63.91 & 64.11 & 64.11 & 64.04 \\
        \noalign{\smallskip}
        NND & \underline{5.29} & 5.50 & 5.50 & 5.43
        & \underline{22.53} & 25.98 & 25.71 & 24.74
        & 39.74 & 43.18	& 42.31 & 41.74 \\
        \noalign{\smallskip}
        PCA & 5.30 & 5.66 & 5.49 & 5.48
        & 34.44 & 35.78 & 31.25 & 33.82
        & 43.02 & 46.64 & 45.78	& 45.15 \\
        \noalign{\smallskip}
        Improved Wavelet & 9.67 & 10.51 & 7.59 & 9.26 & 43.47 & 43.48 & 39.69 & 42.21 & 36.54 & 37.36 & 34.05 & 35.98  \\\noalign{\smallskip}
        AEF & 5.30 & 5.34 & 5.14 & 5.26 & 23.87 & 23.11 & 20.58 & 22.52 & 39.07 & 40.60 & 36.82 & 38.83  \\\noalign{\smallskip}
        CCFL & 6.60 & 6.45 & 7.35 & 6.80 & 28.89 & 32.91 & 37.81 & 33.20 & \underline{29.50} & \textbf{29.53} & \textbf{26.85} & \textbf{28.62} \\\noalign{\smallskip}
        Ours & \textbf{4.25} & \textbf{4.64} & \underline{4.56} & \textbf{4.48}
        & \textbf{16.43} & \textbf{17.72} & \underline{19.59} & \textbf{17.91}
        & \textbf{29.24} & \underline{31.72} & 30.58 & \underline{30.52} \\
	\noalign{\smallskip}\hline
	\end{tabular}}}
\end{table*}

To summarize, a comprehensive analysis of multiple metrics is needed to judge their performance due to the complexity and particularity of remote sensing data. 
The wavelet fusion method has excellent non-redundancy, reconstruction, and detailed texture retention ability. However, the obtained images are too close to the multispectral images, which is unsuitable for our requirement of optimizing the intermediate representation. 
The fusion results of HSV and PCA are also not satisfactory.
HSV replaces the converted luminance value band with SAR images to preserve the details of SAR images. Because of the differences in spectral reflectance curves in different bands, HSV will inevitably produce spectral degradation, distortion, and noise during component replacement, leading to terrible fusion results.
The PCA fusion algorithm shall divide multispectral images into different levels of principal components, and simply replace the first one with the SAR images. It applies to all bands of the multispectral images. However, because of simply removing the first principal component, PCA will lose some information reflecting spectral characteristics, which results in severe spectral distortion. It does not fully consider the features from different levels of principal components.
NND and GS show similar but slightly lower performance than our proposed method, hindered by some speckle noise and distortion.
The two CDL-based solutions (AEF and CCFL) both achieve outstanding performance and similar visual results to our solution, but still perform lower than our solution based on the comprehensive comparisons.
Our proposed method attains superior qualitative and quantitative performance, realizing the best performance in the degree of spectral distortion, MSE, NIQE, and BRISQUE, and the second best in the correlation coefficient and PIQE. Visual evaluation of our proposed method also indicates excellent noise removal and distinctive texture detail preservation. These experimental results demonstrate the superior performance of our proposed pseudo-color fusion method.

\section{Conclusion}
\label{sec:5}

This paper proposes a novel pseudo-color fusion algorithm for remote sensing applications based on CDL and Brovey pre-processing. The method performs further processing on the multispectral image and the pre-processed Brovey image, capturing the relevant information of both images with CDL and reconstructing the images with the reconstruction errors of joint sparse representation. The results show that the fused images trade off the linear relationship between SAR and multispectral images, optimizing their intermediate representation. Our methodology shows exceptional performance in all objective metrics and subjective visual evaluation. 
However, a significant limitation of our method lies in its application to different modalities. Our approach primarily focuses on the fusion of single-channel and three-channel images, achieving high spatial and spectral resolution in the fused image. Therefore, in similar applications such as PEI and MRI, visible and infrared images, where the combination of structural details from single-channel images and spectral information from multi-channel images is required, there are similarities to SAR and multispectral image fusion. Hence, our method holds potential in these areas. However, its performance may not be satisfactory when applied to multi-focus images. Future work will involve (i) validating our method on image fusion tasks involving different modalities, (ii) exploring the potential of combining our CDL and sparse representation method with existing fusion methodologies to optimize performance, and (iii) assessing the applicability of our fusion methodology in downstream remote sensing tasks, such as remote sensing object segmentation and detection.

\bibliographystyle{IEEEtran}
\bibliography{reference}

\begin{IEEEbiography}
[{\includegraphics[width=1in,height=1.25in,clip,keepaspectratio]{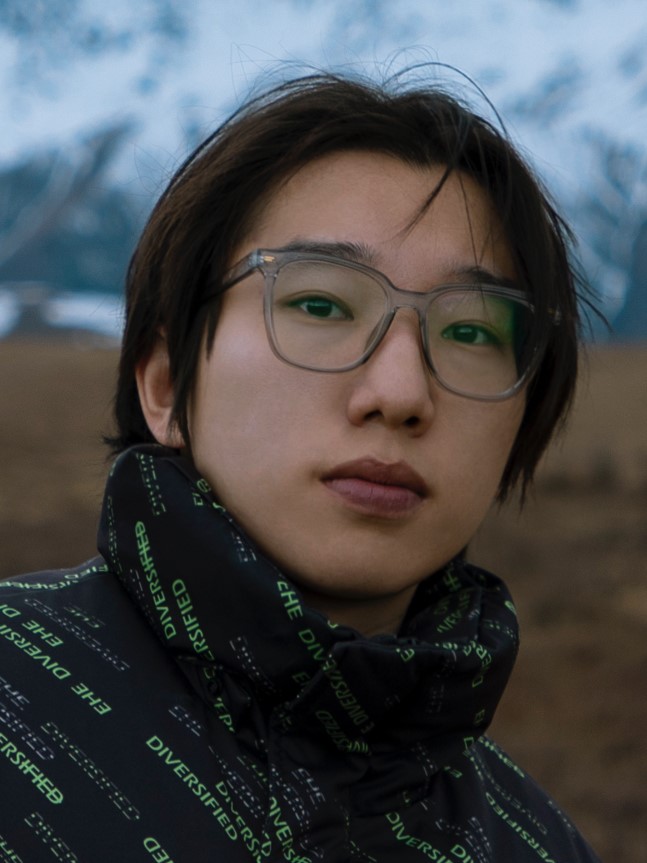}}]{Long Bai} (Graduate Student Member, IEEE)
received the B.S. degree in Opto-Electronics Information Science and Engineering from the Beijing Institute of Technology (BIT) at Beijing, China, in 2021. He is currently pursuing the Ph.D. degree with the Department of Electronic Engineering, The Chinese University of Hong Kong (CUHK), Hong Kong. He is a recipient of the CUHK Vice-Chancellor's Ph.D. Scholarship Scheme (2021), the IEEE RAS Travel Grants Award (2023), the Best Poster Award in ICRA workshop on surgical robotics (2023), and the ICBIR Best Student Paper Award (2023). His current research interests include robotics perception, capsule endoscopy, surgical scene understanding, continual learning, and human-robot interaction.
\end{IEEEbiography}

\begin{IEEEbiography}
[{\includegraphics[width=1in,height=1.25in,clip,keepaspectratio]{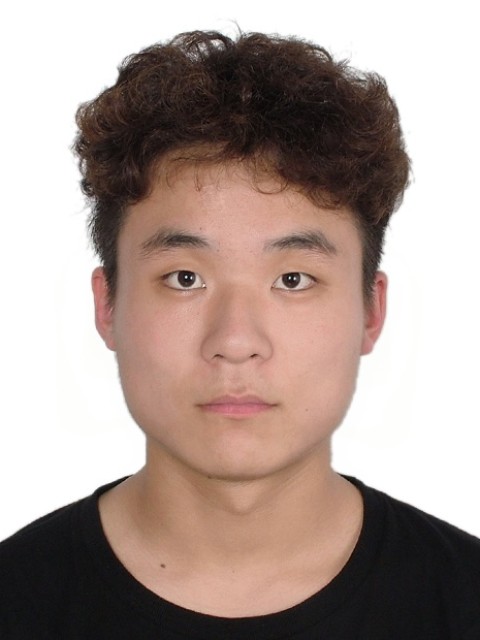}}]{Shilong Yao}
received the B.E. degree in Mechanical Engineering from the Southern University of Science and Technology, Shenzhen, China, in 2021. He is currently pursuing the Ph.D. degree with the Department of Electrical Engineering, City University of Hong Kong (CityU), Hong Kong. 

His current research interests include surgical robotic systems and machine learning in medical robotics.
\end{IEEEbiography}

\begin{IEEEbiography}
[{\includegraphics[width=1in,height=1.25in,clip,keepaspectratio]{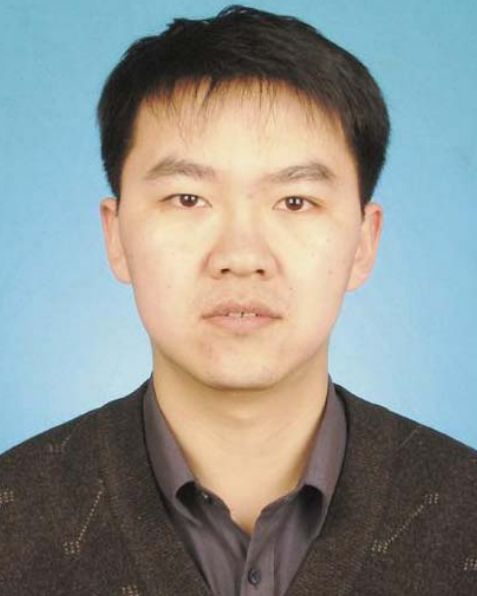}}]{Kun Gao}
received the B.A. degree in Electrical Engineering and the Ph.D. degree in Instrument Science and Engineering from Zhejiang University, China, in 1995 and 2002, respectively. From 2002 to 2004, he was a Postdoctoral Fellow with Tsinghua University, Beijing, China. Since 2005, he has been with the Beijing Institute of Technology, Beijing, with a focus on infrared technology and real-time image processing. He is a member of the Optical Society of China.

\end{IEEEbiography}

\begin{IEEEbiography}
[{\includegraphics[width=1in,height=1.25in,clip,keepaspectratio]{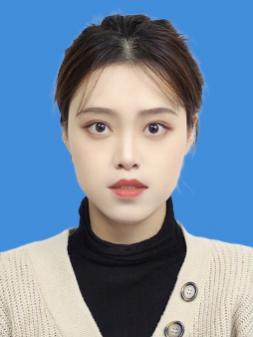}}]{Yanjun Huang}
received the B.S. degree in Opto-Electronics Information Science and Engineering from the Beijing Institute of Technology (BIT) at Beijing, China, in 2021. She is currently pursuing the M.S degree with the School of Optics and Photonics, Beijing Institute of Technology (BIT) at Beijing, China.

Her current research interests include Synthetic Aperture Radar (SAR) image processing.
\end{IEEEbiography}

\begin{IEEEbiography}
[{\includegraphics[width=1in,height=1.25in,clip,keepaspectratio]{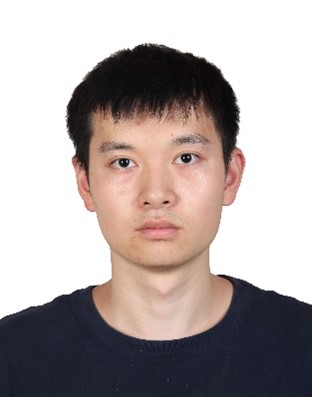}}]{Ruijie Tang}
received the B.S. degree in Mechanical Engineering from the South China University of Technology, Guangzhou, China and University of Edinburgh, Edinburgh, the United Kingdom, in 2018, and the M.S degree in Advanced Aeronautical Engineering at Imperial College London, London, the United Kingdom in 2019. He is currently pursuing the Ph.D. degree in Electronic Engineering, The Chinese University of Hong Kong. 
His current research interests include soft robots, force sensor design, magnetic actuated robots with application in medical field. 

\end{IEEEbiography}

\begin{IEEEbiography}
[{\includegraphics[width=1in,height=1.25in,clip,keepaspectratio]{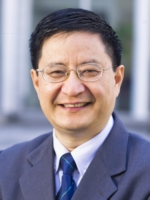}}]{Hong Yan}
(Fellow, IEEE) received the Ph.D. degree from Yale University. He was Professor of Imaging Science with the University of Sydney and currently is Wong Chun Hong Professor of Data Engineering and Chair Professor of Computer Engineering with City University of Hong Kong. His research interests include image processing, pattern recognition and bioinformatics. He has authored or coauthored over 600 journal and conference papers in these areas. He is an IAPR Fellow, a member of European Academy of Sciences and Arts, and a Fellow of the US National Academy of Inventors. He received the 2016 Norbert Wiener Award for contributions to image and biomolecular pattern recognition techniques from the IEEE SMC Society.
\end{IEEEbiography}

\begin{IEEEbiography}
[{\includegraphics[width=1in,height=1.25in,clip,keepaspectratio]{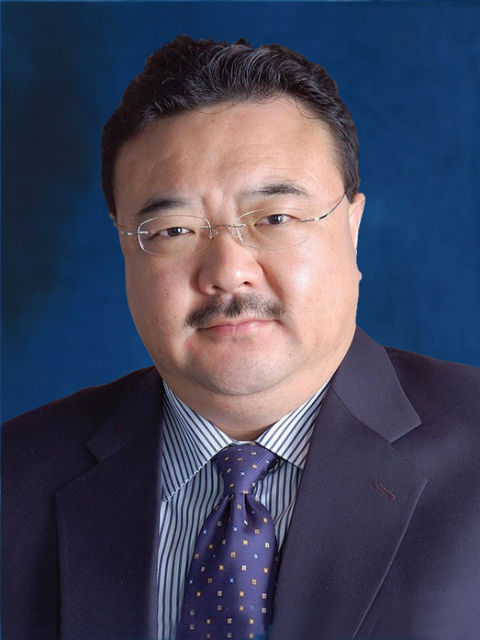}}]{Max Q.-H. Meng}
(Fellow, IEEE) received the Ph.D. degree in electrical and computer engineering from the University of Victoria, Victoria, BC, Canada, in 1992.

He is currently a Chair Professor and the Head of the Department of Electronic and Electrical Engineering with the Southern University of Science and Technology, Shenzhen, China, on leave from the Department of Electronic Engineering, The Chinese University of Hong Kong, Hong Kong. He joined the Chinese University of Hong Kong in 2001 as a Professor and later the Chairman of Department of Electronic Engineering. He was with the Department of Electrical and Computer Engineering, University of Alberta, Edmonton, AB, Canada, where he was the Director of the Advanced Robotics and Teleoperation Lab and held the positions of Assistant Professor in 1994, an Associate Professor in 1998, and a Professor in 2000. He has authored or coauthored more than 750 journal and conference papers and book chapters and led more than 60 funded research projects to completion as a Principal Investigator. His research interests include robotics, perception, and intelligence.

Prof. Meng is a fellow of the Hong Kong Institution of Engineers and an Academician of the Canadian Academy of Engineering and an Elected Member of the AdCom of IEEE RAS for two terms. He was a recipient of the IEEE Millennium Medal. He is the General Chair or Program Chair of many international conferences, including the General Chair of IROS 2005 and ICRA 2021, respectively. He served as an Associate VP for Conferences of the IEEE Robotics and Automation Society from 2004 to 2007 and the Co-Chair of the Fellow Evaluation Committee. He is the Editor-in-Chief and with the editorial board of a number of international journals, including the Editor-in-Chief of the Elsevier Journal of \emph{Biomimetic Intelligence and Robotics}.
\end{IEEEbiography}

\begin{IEEEbiography}
[{\includegraphics[width=1in,height=1.25in,clip,keepaspectratio]{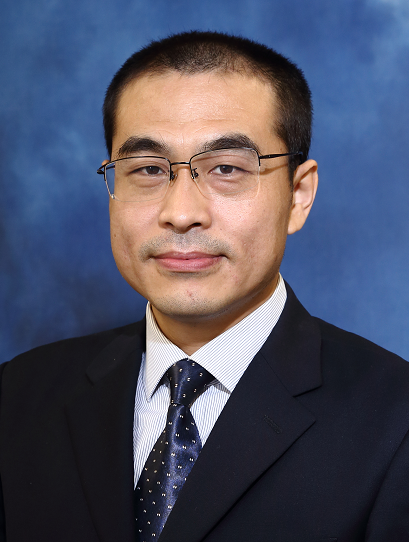}}]{Hongliang Ren}
(Senior Member, IEEE) received the Ph.D. degree in electronic engineering (specialized in biomedical engineering) from The Chinese University of Hong Kong (CUHK) in 2008. He has navigated his academic journey through CUHK, Johns Hopkins University, Children’s Hospital Boston, Harvard Medical School, Children’s National Medical Center, USA, and National University of Singapore (NUS). He is currently an Associate Professor with the Department of Electronic Engineering, CUHK, and an Adjunct Associate Professor with the Department of Biomedical Engineering, NUS. His research interests include biorobotics, intelligent control, medical mechatronics, soft continuum robots, soft sensors, and multisensory learning in medical robotics.

He was a recipient of the NUS Young Investigator Award, the Engineering Young Researcher Award, the IAMBE Early Career Award 2018, the Interstellar Early Career Investigator Award 2018, the ICBHI Young Investigator Award 2019, and Health Longevity Catalyst Award 2022 by NAM \& RGC. He serves as an Associate Editor for \emph{IEEE Transactions on Automation Science and Engineering} and \emph{Medical \& Biological Engineering \& Computing (MBEC)}.
\end{IEEEbiography}

\end{document}